# Observation of bright and dark exciton transitions in monolayer MoSe$_2$ by photocurrent spectroscopy

*Jorge Quereda\*, Talieh S. Ghiasi, Feitze A. van Zwol, Caspar H. van der Wal, Bart J. van Wees*

Physics of Nanodevices, Faculty of Science and Engineering, Zernike Institute for Advanced Materials, University of Groningen, Groningen, The Netherlands

\* e-mail: j.quereda.bernabeu@rug.nl

**Abstract:**

We investigate the excitonic transitions in single- and few-layer MoSe$_2$ phototransistors by photocurrent spectroscopy. The measured spectral profiles show a well-defined peak at the optically active (bright) A$^0$ exciton resonance. More interestingly, when a gate voltage is applied to the MoSe$_2$ to bring its Fermi level near the bottom of the conduction band, another prominent peak emerges at an energy 30 meV above the A$^0$ exciton. We attribute this second peak to a gate-induced activation of the spin-forbidden dark exciton transition, A$_D^0$. Additionally, we evaluate the thickness-dependent optical bandgap of the fabricated MoSe$_2$ crystals by characterizing their absorption edge.

## 1. Introduction

Atomically thin transition metal dichalcogenides (TMDCs) [1–3] are very attractive materials for two-dimensional (2D) spintronics, since their large spin-orbit splitting and coupled spin and valley physics allow for the optical generation of spin currents in these crystals.[4–8] Several groups have recently explored the possibility of using light of a specific polarization and wavelength to selectively populate spin-resolved levels in monolayer TMDCs,[9–15] which opens new possibilities for the design of opto-spintronic devices.

In order to fully advance and exploit the optical generation of spin-polarized carriers it is necessary to test and develop the understanding of optically active excitonic states of TMDCs and their contribution to electronic and spin transport. Photocurrent spectroscopy is a powerful technique that allows to access the excitonic states of nanoscaled semiconductor materials.[5,16–19] Further, when used in combination with ferromagnetic electrodes, it allows to characterize the spin of photogenerated charge carriers. [14] However, only few efforts towards applying this technique to two-dimensional TMDCs have been reported so far.

In this work we investigate the near-infrared photocurrent spectra of single- and few-layer MoSe$_2$ phototransistors and their dependence on the crystal Fermi energy. The resulting photocurrent spectra allow to directly observe the bright A$^0$ exciton resonance of atomically thin MoSe$_2$, as well as its dependence on the flake thickness. More interestingly, when a gate voltage is used to increase the carrier population in the conduction band of the MoSe$_2$ crystals, a second prominent peak appears at an energy ~30 meV higher than that of the bright A$^0$ exciton. Our analysis shows that this peak is caused by a resonant excitation of the dark exciton state, A$_D^0$, of the MoSe$_2$ crystal. [16,17] To the best of our knowledge, this work is the first report of such effect in TMDCs, opening the possibility of controlling the dark exciton population in atomically thin semiconductors by a combination of gate-induced doping and optical pumping.

Observation of bright and dark exciton transitions in monolayer MoSe$_2$ by photocurrent spectroscopy

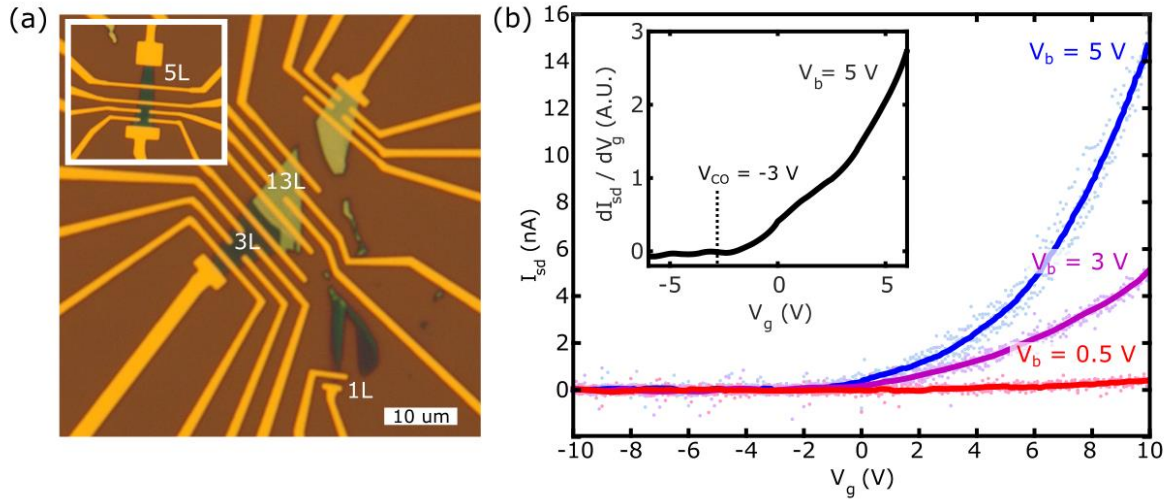

**Fig. 1.** (a) Optical micrograph of the fabricated phototransistors based on exfoliated MoSe$_2$. The crystal thicknesses range from monolayer (1L) to 13 layers. All the devices are fabricated on a single SiO$_2$/Si substrate. The Ti/Au contacts are fabricated on top of the MoSe$_2$ crystals, patterned by e-beam lithography. (b) Transfer characteristics of the monolayer device shown in panel (a) under three different bias voltages: $V_b$ = 0.5 V, 3 V and 5 V. Inset: Derivative of the source-drain current as a function of the gate voltage for $V_b$ = 5 V. The gate voltage at which the source-drain current starts to increase, $V_{CO}$ ~ -3 V, can be identified by the sudden change in the slope of the derivative.

**2. Sample fabrication and electrical characterization**

**Fig. 1a** shows the fabricated MoSe$_2$ field-effect transistors. In an effort to reduce the inhomogeneity between the different devices, all the studied phototransistors are fabricated simultaneously on a single substrate. We first exfoliate the crystals onto a SiO$_2$/Si substrate with an oxide thickness of 300 nm. Then, we use optical microscopy to identify atomically thin flakes, and measure their thickness by atomic force microscopy, as shown in the Supporting Information. The electrical contacts are fabricated on top of the selected crystals using standard e-beam lithography (EBL) and e-beam evaporation of Ti (5 nm) / Au (55 nm). The resulting sample contains four MoSe$_2$ phototransistors with flake thicknesses of 1, 3, 5 and ~13 atomic layers.

**Fig. 1b** shows the transfer characteristics of the monolayer MoSe$_2$ phototransistor under three different bias voltages. The device shows a clear n-doping character, and the MoSe$_2$ channel starts to become open for charge transport at a gate voltage of $V_{CO}$ ~ -3 V, as can be clearly identified by a sudden slope change in the derivative of the transfer characteristic (see inset of Fig. 1b). A similar n-type behavior is also found in the multilayer devices, with $V_{CO}$ ranging between -4 V and -40 V.

## 3. Optical response and photocurrent spectroscopy

Next, we investigate the photocurrent spectrum of the MoSe$_2$ devices. **Fig. 2a** shows schematically the experimental setup. First, the sample is loaded in a vacuum chamber to prevent the adsorption of impurities to the MoSe$_2$ surface. Then, we use a tunable infrared continuous-wave laser (See Supporting Information for details) to illuminate the sample while registering the current under a constant bias voltage ($V_b$ = 5 V). To prevent the system from heating during the measurement, the laser power density was kept below 10 pW/μm$^2$. Additionally, we use a shutter to rapidly switch on the illumination just before the current is measured and then off again after the measurement is finished, so the MoSe$_2$ crystal is only exposed to the laser source for a time interval of 0.1 s. Then, we keep the system in dark for 5 seconds before the next measurement is performed. These precautions remove the presence of slow drifts in the photocurrent, which probably occur due to optically-induced shifts in the intrinsic doping (consistent with observed shifts in the gate-voltage dependence when first bringing the device in a vacuum environment) and bulk heating effects.

To achieve a uniform illumination power density along the whole sample the laser beam is expanded up to a diameter of 1 cm, much larger that the studied devices (see Supp. Info. section 1). Then, the beam is collimated to ensure that the illumination is perpendicular to the MoSe$_2$ surface.

The polarization and helicity of the illumination source can be controlled using $\lambda/2$ and $\lambda/4$ waveplates, as shown schematically in Fig 2a. However, we did not detect any polarization or helicity dependence in the measured spectra, consistent with the expected behavior when driving the transitions well below saturation (see also Supp. Info. section 2). For consistency, unless otherwise stated, all the spectra shown here are acquired using linearly polarized light.

Given the n-type character of our devices (see Figure 1b), it is necessary to apply a negative gate voltage to compensate the intrinsic doping. Further discussion about the effect of gate voltage in the measured photocurrent spectra can be found below. **Fig. 2b** shows a photocurrent spectral profile from the monolayer MoSe$_2$ device while applying a gate voltage $V_g$ = -6 V. The photocurrent is obtained as the difference between the current measured when the device is exposed to light (bright current) and the one measured in dark immediately before exposure (dark current). Further details regarding the measurement process can be found in the Supporting Information.

The resulting spectrum shows a prominent peak in the photocurrent at 1.59 eV, accurately matching with the A$^0$ exciton resonance reported in literature from photoluminescence, absorption and micro-reflectance spectroscopy measurements in monolayer MoSe$_2$ (for the specific case of room temperature and a SiO$_2$ (300 nm) / Si substrate the A$^0$ exciton peak is found at 1.58±0.01 eV [20–22]). When the photon energy reaches 1.7 eV, the photocurrent starts to increase again with the photon energy. This is probably due to the proximity of the B$^0$ exciton resonance, which is expected to be at 1.79 eV (200 meV above the A$^0$ exciton). However, the spectral range of the used laser did not allow to reach energies above 1.76 eV, and therefore we cannot detect the B$^0$ exciton peak. For photon energies below 1.5 eV we do not observe any photoresponse from the device, and the spectral profile remains featureless, pointing to a low density of midgap states due to the high quality of the MoSe$_2$ crystal.

It must be noted that, in order to produce photocurrent, a neutral exciton must first dissociate into an unbound electron-hole pair. It has been reported that, in the case of monolayer TMDC phototransistors, this exciton dissociation can be caused by the strong electric fields that arise in the semiconductor crystal near the metallic electrodes when a bias voltage is applied.[5] In our experiments we are able to observe the exciton resonances at bias voltages as low as $V_b = 0.5$ V, showing that the higher value of the bias voltage $V_b = 5$ V used in most of our measurements is not essential for the observed features. At bias voltages below 0.5 V, the photocurrent signal drops below the noise level.

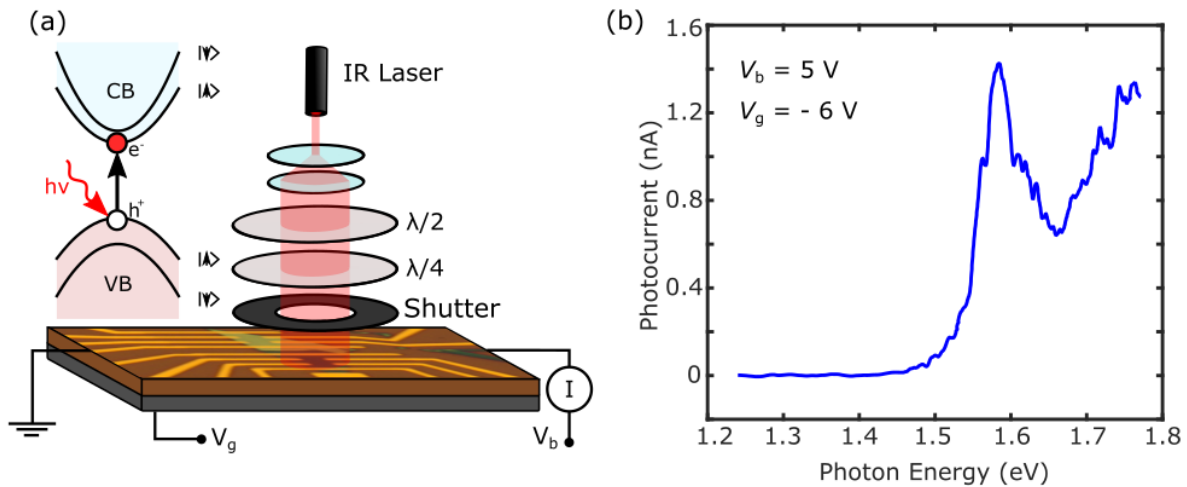

**Fig. 2.** (a) Schematic of the experimental setup. The whole sample is uniformly illuminated using a tunable infrared laser source. A halfwave plate and a quarterwave plate are used to control respectively the polarization and helicity of the illumination source. When the photon energy of the laser is in resonance with an exciton transition of the crystal, a peak is observed in the measured photocurrent. Inset: Schematic of the optical pumping mechanism at one of the valley points of the MoSe$_2$ reciprocal lattice. When the photon energy of the illumination source ($hv$) is above the MoSe$_2$ optical bandgap (accounting for exciton binding energies), excitons and electron-hole pairs can be generated. (b) Photocurrent spectroscopy profile measured at a monolayer MoSe$_2$ phototransistor. The $A^0$ exciton peak is observed at 1.59 eV.

**4. Gate-induced activation of the dark exciton transition.**

We now discuss the influence of the gate voltage on the measured spectra. **Fig. 3** shows the photocurrent spectra obtained for monolayer MoSe$_2$ under different gate voltages, between $V_g$ = -6 V and $V_g$ = 10 V. When the gate voltage is increased above $V_g$ = -3 V (and the conduction-band carrier concentration reaches values that allow for charge transport, as shown in Fig. 1b) a second well-defined peak appears at a photon energy of 1.62 eV, 30 meV above the energy of the $A^0$ exciton. The energy difference between these two peaks closely matches the reported conduction band spin-orbit splitting of MoSe$_2$[23], indicating that this second peak corresponds to the spin-forbidden dark exciton transition $A_D^0$ (further discussed below), schematically depicted in the band diagram shown in Figure 3. Note that this peak cannot be associated with the trion states $A^{+/-}$,[22] which occur 30 meV below the $A^0$ exciton peak (indeed observed in our spectra as a small but persistent feature at 1.56 eV). The spectra measured for gate voltages $V_g$ > -3 V also show a prominent peak at 1.75 eV, 160 meV above the $A^0$ exciton. This peak could be related with the spin-forbidden $B_D^0$ transition (which is lower in energy than $B^0$, see inset Fig. 3) or the trion states $B^{+/-}$, whose population is expected to increase with the doping of the MoSe$_2$ crystal. [22] These two cases are hard to distinguish, since both $B_D^0$ and $B^{+/-}$ appear at similar energies, ~ 30 meV below the $B^0$ resonance (see inset in Fig. 3). When the gate voltage is further increased up to $V_g$ = 10 V the peak associated to the $A_D^0$ transition becomes smaller compared with the $A^0$ peak. This could be due to an increased population of the $A_D^{+/-}$ trion states, expected to appear at an energy similar to that of the $A^0$ peak.

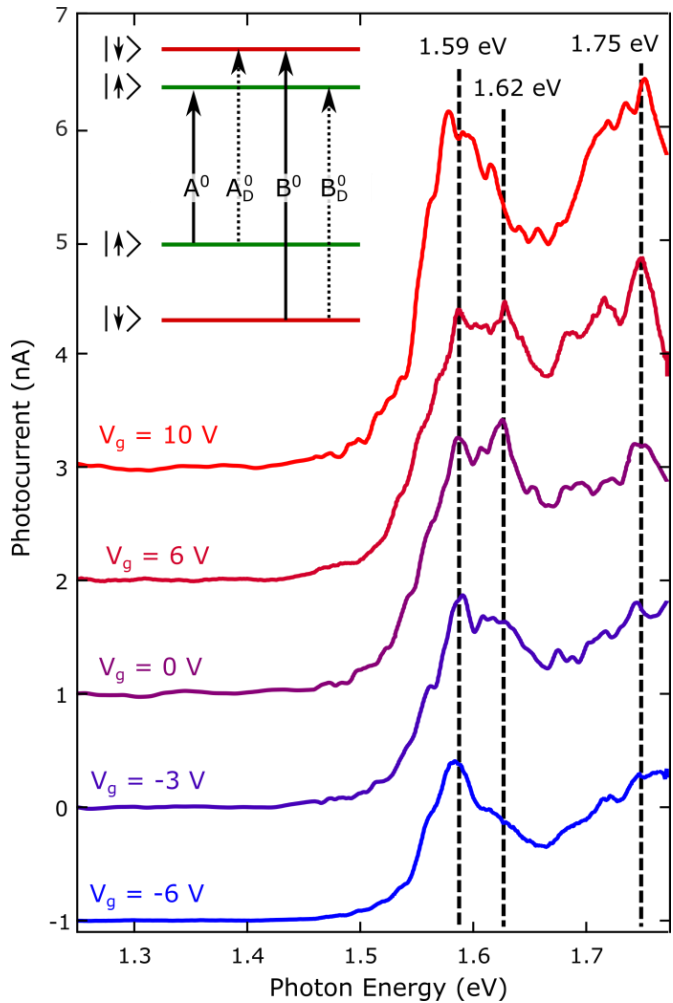

**Fig. 3.** Photocurrent spectra of monolayer MoSe$_2$ under different gate voltages between $V_g$ = -6V and $V_g$ = +10V. The bias voltage is kept at $V_b$ = +5V. Three well defined peaks are observed in the spectra at 1.59 eV, 1.62 eV and 1.75 eV. As discussed in the main text, we associate this peaks to the $A^0$, $A_D^0$ and $B^{+/-}$ excitonic states. Inset: Diagram illustrating the possible exciton transitions at the $K$ valley of MoSe$_2$. The trion transitions $A^{+/-}$ and $B^{+/-}$ (not depicted here) are expected to occur at an energy ~ 30 meV below the $A^0$ and $B^0$ transitions respectively.

The observation of a strong peak for the $A_D^0$ transition is intriguing, since it requires a spin flip and, therefore, should not be optically active. However, similar gate-induced activation of dark exciton transitions has been recently observed in the photocurrent spectra of carbon nanotubes. [16,17,24] Remarkably, such gate-induced dark exciton peaks are much more prominently observed in photocurrent spectroscopy than in photoluminescence, since the exciton dissociation process is faster than the radiative recombination required for photoluminescence.[24] This could explain the absence of previous observations of a gate-induced opening of the $A_D^0$ transition in the photoluminescence spectra of TMDC.

For alternative interpretations for the $A_D^0$ peak, we consider but rule out the following possibilities: For these lowest transitions, a phonon-side peak (which can enhance parity-forbidden transitions in nanotubes[17]) is not expected, since it would require very high phonon energies or coupling, such that it can mediate spin or valley scattering. Surface plasmon-polaritons can enhance spin-forbidden transitions in TMDCs[25], but only occur when applying optical fields with a significant in-plane component (and have only been observed at 4 K). Cavity-polaritons can indeed give splittings of the A$^0$ peak of 30 meV[26], but this requires an optical cavity that gives strong coupling. While our material is a layered structure, the interface reflectivities are too low for the formation of such a cavity. Also, our set of two peaks is not observed as a splitting around the usual position of the A$^0$ resonance (A$^0$ does not shift away from the value 1.59 eV). Given our detection technique, modified transitions may occur due to band bending in the depletion regions near the metal contacts. However, our A$^0$ peak occurs exactly where it was observed for photoluminescence, and the peaks do not shift or broaden with bias or gate voltage, which makes this interpretation implausible. Finally, one could consider a role for defects or impurities. However, due to the nonlocal character of the spectroscopy technique (the whole device is uniformly illuminated), the effect of local defects in the resulting spectra is very limited, as further supported by the lack of defect-induced spectral features below the absorption edge. We thus interpret the peak at 1.62 eV as the A$_D^0$ transition.

Slobodeniuk and Basko[27] recently suggested two mechanisms that can open the spin-forbidden A$_D^0$ transition in TMDCs. First, in the presence of an out-of-plane electric field, a Bychkov-Rashba coupling term appears in the TMDC Hamiltonian, leading to the opening of the transition (also discussed in refs. [28,29]). However, they estimate that by this mechanism the A$_D^0$ transition remains about a factor 1000 weaker than A$^0$, which is not strong enough for explaining the prominent A$_D^0$ feature in our data. Alternatively, they propose that the presence of charge carriers in the conduction band can lead to an electron-induced intervalley dark exciton transition (which would require a large electron momentum kick without a spin flip). This suggestion is consistent with our observation that the A$_D^0$ transition opens when the gate induces an increased charge carrier population in the conduction band. However, a detailed analysis of such a mechanism has not been described yet.

**5. Photocurrent spectra of multilayer devices and thickness-dependent optical bandgap.**

Finally, we investigate how our photocurrent-spectroscopy technique can be applied to few-layer MoSe$_2$ devices. Due to broadening of peaks for devices with thicker layers we could not identify the spectroscopy peaks as well as for the monolayer device, but could study the evolution of the optical bandgap with the number of MoSe$_2$ layers. We measure the photocurrent spectra of the fabricated devices, with MoSe$_2$ thicknesses ranging from a monolayer to ~13 atomic layers. **Fig. 4a** shows the thickness-dependent photocurrent spectra of the fabricated MoSe$_2$ devices at a bias voltage $V_b$ = 5 V. The gate voltage was selected in each case to compensate the intrinsic doping of the device. As observed in the figure, when the number of layers is increased, the absorption edge is shifted towards lower energies, due to the thickness-dependent bandgap of MoSe$_2$. Also, the width of the observed spectral features becomes broader for thicker crystals as a result of their increased inhomogeneity, and a higher background photocurrent appears for low photon energies, pointing to a reduced crystallinity and higher density of midgap states. Although the spectra of the multilayer crystals could also present a peak corresponding to the A$_D^0$ excitonic transition, the higher broadening of the spectral features makes it difficult to unambiguously identify this peak.

Observation of bright and dark exciton transitions in monolayer MoSe$_2$ by photocurrent spectroscopy

The measured spectra can be used to extract information about the optical bandgap of the MoSe$_2$ crystals. According to Tauc et al, [30] near the absorption edge, the absorption coefficient of a semiconductor is given by

$$\alpha(\lambda) = A \frac{\left(h\nu - E_g^{opt}\right)^n}{h\nu}, \quad (1)$$

where $A$ is the material-dependent effective Richardson constant, $E_g^{opt}$ is the optical bandgap, $h\nu$ is the photon energy and $n = 1/2$ for direct transitions. Assuming that near the absorption edge the photocurrent $I_{pc}$ is proportional to the absorption coefficient $\alpha$ we get [19]

$$\left(I_{pc}h\nu\right)^2 \propto \left(h\nu - E_g^{opt}\right). \quad (2)$$

The usual method for determining the value of $E_g^{opt}$ (referred in literature as Tauc extrapolation) [19,31,32] involves representing $(I_{pc}h\nu)^2$ as a function of the photon energy and then fitting the absorption edge to a linear function (shown in Fig. 4b). Then, according to equation (2), the intersection of this linear fit with the horizontal axis gives the value of the optical bandgap. The inset in Figure 4b shows the estimated optical bandgap as a function of the number of layers. We observe a ~80 meV decrease of the optical bandgap, from 1.54 eV in the monolayer to 1.46 eV in the thicker crystals. This results are in good agreement with the thickness-dependent optical bandgap reported in literature for atomically thin MoSe$_2$.[33]

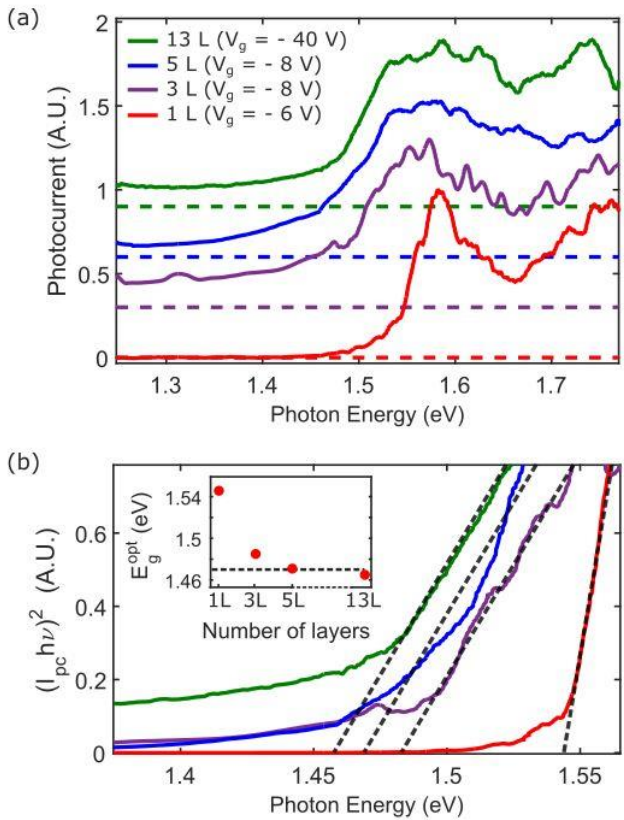

**Fig. 4.** (a) Thickness-dependent photocurrent spectra of MoSe$_2$. Four different devices are shown, with thicknesses ranging between monolayer and 13 layers. (b) Plot of $(I_{pc}h\nu)^2$ as a function of the photon energy. The black dashed lines are the Tauc extrapolations of the absorption edge. Inset: Thickness-dependent optical bandgap estimated from the Tauc extrapolations (red circles) and estimated optical bandgap for bulk MoSe$_2$ [33] (dashed line).

## 5. Conclusions

In conclusion, we have investigated the exciton states of atomically thin MoSe$_2$ by photocurrent spectroscopy. The $A^0$ exciton optical transition, occurring at a photon energy of 1.59 eV was clearly identified in the photocurrent spectra acquired in the monolayer MoSe$_2$ device. Furthermore, using a gate voltage to tune the Fermi energy of the monolayer MoSe$_2$ crystal around the edge of the conduction band, we are able to turn on and off the spin-forbidden dark exciton optical transition $A_D^0$. The photocurrent spectra acquired in multilayer MoSe$_2$ devices also allowed to characterize the direct bandgap of MoSe$_2$ and its dependence on the flake thickness.

The results shown here open new possibilities for optically addressing the dark exciton states of two-dimensional transition metal dichalcogenides, which could be of great interest for the design of optoelectronic and spintronic devices.

**Author Contributions**

The manuscript was written through contributions of all authors. All authors have given approval to the final version of the manuscript.

**Funding Sources**

This work was financed by the Foundation for Fundamental Research on Matter (FOM), which is part of the Netherlands Organisation for Scientific Research (NWO), and supported by the Zernike Institute for Advanced Materials.

Observation of bright and dark exciton transitions in monolayer MoSe$_2$ by photocurrent spectroscopy

**Acknowledgments**

We thank Tom Bosma and Jakko de Jong for contributions to the laser control system. We thank H. M. de Roosz, J.G. Holstein, H. Adema and T.J. Schouten for technical assistance.

# Supporting information

**Table of contents:**

Observation of bright and dark exciton transitions in monolayer $MoSe_2$ by photocurrent spectroscopy

**Characterization of the illumination intensity**

The tunable laser source used for this work was a Solstis 2000 SRX-XF infrared Ti:Sapphire continuous wave laser from M-squared, with a wavelength range covering the spectral window between 700 and 1000 nm and a narrow linewidth of 1MHz. The Solstis infrared laser is driven by a Coherent Verdi 18W pump laser with a wavelength of 523nm. The output wavelength is registered using an Angstrom WS7 wavelength meter, connected to the Solstis laser by a single mode fiber. A power attenuation unit has been placed in the optical path to lower the output power of the laser from 3 W to around 15mW, to prevent damage to the sample.

To ensure a uniform distribution of the illumination intensity along our devices we telescope the laser beam using two lenses with different focal distances. Figure S1 shows the spatial distribution of the spot power density. The maximum value of the gradient of the power density is 50 mm$^{-1}$. The typical size of our devices is of ~1 μm. Consequently, even assuming the worst possible alignment between the sample and the laser spot, the power density can only change by 0.25% between the two electrodes.

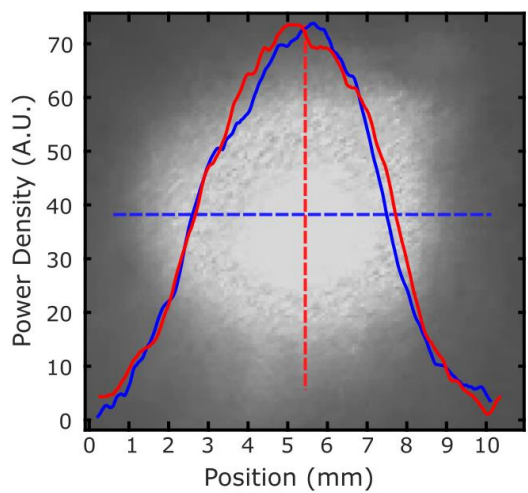

**Fig. S1.** Spatial distribution of the laser spot intensity along the horizontal (blue) and vertical (red) axis. The watermark below the plot is an optical image of the laser spot.

**Power and polarization dependence of the measured photocurrent**

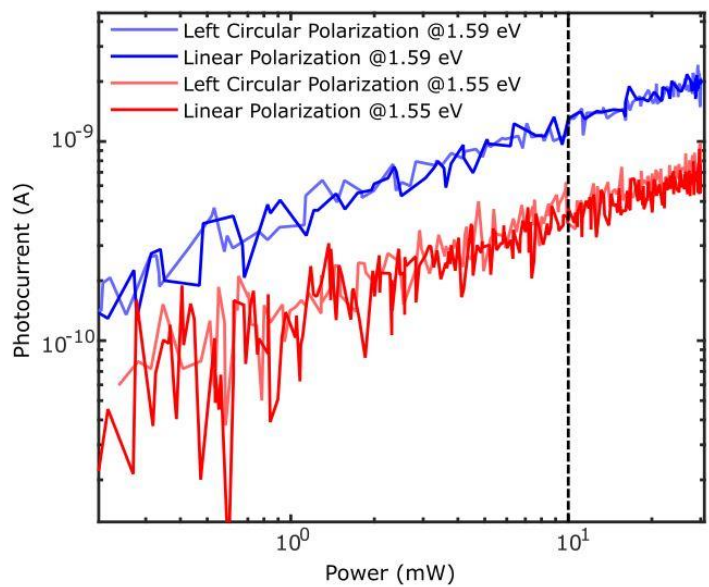

**Fig. S2.** Measured photocurrent for the monolayer device as a function of the laser power for two different polarizations (linear and left circular) and two different illumination energies: 1.59 eV (on resonance with the A$^0$ excitonic transition) and 1.55 eV (off resonance).

Figure S2 shows a logarithmic plot of the measured photocurrent in the monolayer MoSe$_2$ phototransistor as a function of the illumination power, both for illumination on resonance with the A$^0$ excitonic transition ($h\nu$ = 1.59 eV) and off resonance ($h\nu$ = 1.55 eV). We observe a sublinear dependence of the photocurrent, $I_{PC} \propto P^{0.48}$, for both illumination wavelengths, which indicates that both photoconduction and photogating play a role in the photocurrent generation [S1]. We do not observe any polarization dependence of the photocurrent, as expected when the illumination intensity is far enough below saturation.

**MoSe$_2$ thickness characterization of the MoSe$_2$ crystals**

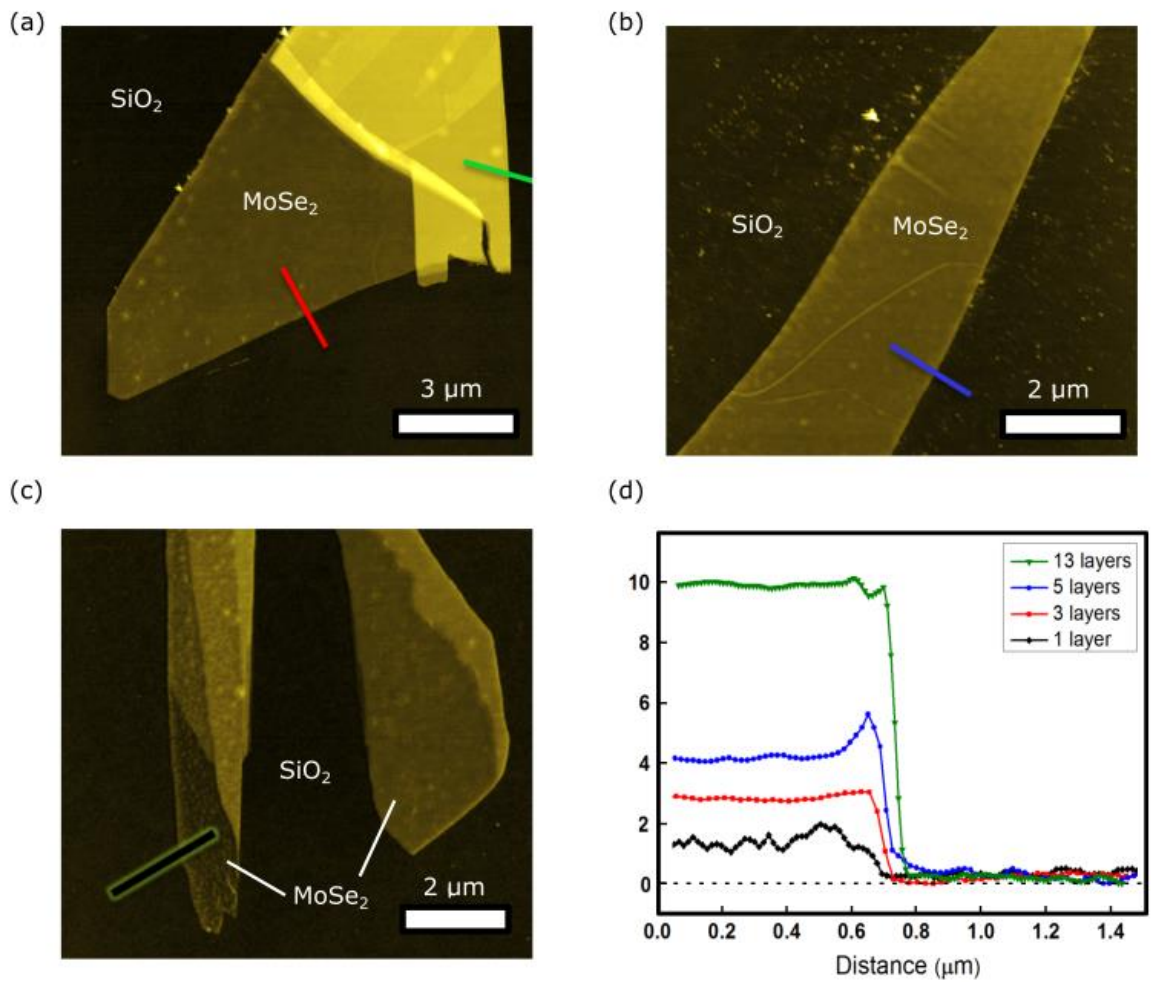

**Fig. S3.** (a-c) AFM topography images of the exfoliated MoSe$_2$ flakes on SiO$_2$ prior to the EBL processing of the electrodes. (d) Scan profiles along the lines marked in the AFM images.

**Photocurrent spectra measurement process**

To acquire a photocurrent spectral profile we start by applying a 5 V bias voltage between the electrodes of the device and then sweep illumination wavelength and measure the photocurrent through the following steps:

1. Set the laser at a specific wavelength
2. Measure the dark drain source current $I_D$
3. Open the shutter to illuminate the device
4. Wait for 0.1 s
5. Measure the bright drain-source current $I_B$
6. Close the shutter
7. Wait for 5 seconds

The photocurrent is then calculated as the difference between the measured bright and dark drain-source current $I_{PC} = I_B - I_D$. To improve the signal-to-noise ratio each measurement is repeated 20 times and the mean value is obtained.

**Estimation of the gate-dependent Fermi energy**

To estimate the variation of the Fermi energy induced by our gate voltage the Si/SiO₂/MoSe₂ stack can be modelled as a parallel plate capacitor, with a capacitance (per unit area) given by

$$C = \frac{\epsilon_r \epsilon_0}{d} \,. \tag{S1}$$

where $\epsilon_r$ is the relative permittivity of SiO₂, $\epsilon_0$ is the permittivity of vacuum, and $d$ is the thickness of the SiO₂ layer. The areal density $n$ of conduction-band electons in the MoSe₂ crystal can then be estimated using

$$n = \frac{C}{e}(V_g - V_0) \,, \tag{S2}$$

where $e$ is the electron charge, $V_g$ is the gate voltage and $V_0$ is an offset voltage that accounts for the fact that the conduction band only starts to be filled when the Fermi energy is close to the conduction band. The calculation of $V_0$ will be discussed below.

The carrier density, $n$, must also must be equal to the total number of occupied states $\Omega(E_F)$ in the conduction band:

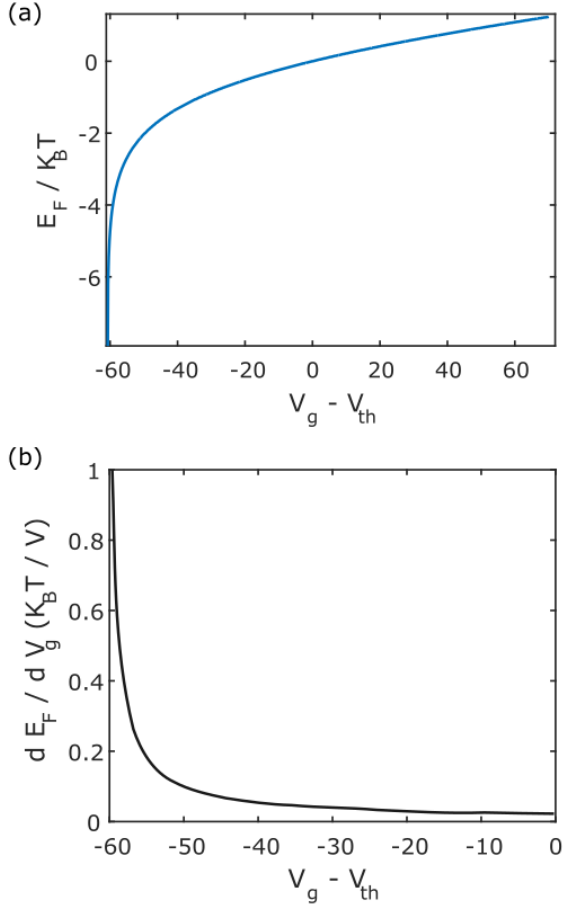

**Fig. S4.** (a) Calculated gate-dependent Fermi energy of the MoSe₂ crystal, in units of the thermal energy $k_B T$. (b) Numeric derivative of the Fermi energy as a function of the gate voltage.

$$\Omega(E_F) = \int_{-\infty}^{\infty} f(E, E_F) \times g_{2D}(E_F)\, dE = n \tag{S3}$$

where $f(E, E_F)$ is the Fermi-Dirac distribution at room temperature and $g_{2D}(E_F)$ is the density of states of a two-dimensional free electron Fermi gas, given by

$$g_{2D}(E_F) = \begin{cases} \dfrac{m_e^*}{2\pi\hbar^2} g_s g_v & \text{if } E \geq 0 \\ 0 & \text{if } E < 0 \end{cases} \tag{S4}$$

where $m_e^*$ is the effective electron mass at the conduction band, $m_e^* = 0.6\, m_e$, $g_s = 1$ is the spin degeneracy and $g_v = 2$ is the valley degeneracy. The energy $E = 0$ is chosen to be at the bottom of the conduction band.

Then, equation S3 yields

Observation of bright and dark exciton transitions in monolayer MoSe$_2$ by photocurrent spectroscopy$$n = \frac{m_e^*}{\pi \hbar^2} \int_0^\infty f(E, E_F) \, dE. \tag{S5}$$

Replacing $n$ by its value from equation S1b and reordering terms we get

$$\int_0^\infty f(E, E_F) \, dE = \frac{C}{e} \times \frac{\pi \hbar^2}{m_e^*} (V_g - V_0). \tag{S6}$$

Finally, we derive the value of $V_0$ from a measured value of the threshold voltage, $V_{th}$, which is the voltage at which the Fermi energy is exactly at the bottom of the conduction band: $E_F(V_{th}) = 0$. From the transfer characteristics of the monolayer device we estimate a value of $V_{th} \sim 30$ V (using how the linear behavior at high $V_g$ extrapolates to zero current). Then,

$$\int_0^\infty f(E, 0) \, dE_F = \frac{C}{e} \times \frac{\pi \hbar^2}{m_e^*} (V_{th} - V_0), \tag{S7a}$$

$$\Rightarrow V_0 = V_{th} - \frac{e \, m_e^*}{C \pi \hbar^2} \int_0^\infty f(E, 0) \, dE. \tag{S7b}$$

Note that, at zero temperature, $f(E,0)$ is a step function and the integral in equation S6b is zero.

Replacing $V_0$ in equation S5 we finally get

$$\int_0^\infty f(E, E_F) \, dE = \frac{C}{e} \times \frac{\pi \hbar^2}{m_e^*} (V_g - V_{th}) + \int_0^\infty f(E, 0) \, dE. \tag{S8}$$

This equation must be solved numerically to obtain $E_F(V_g)$.

Fig. S4a shows the calculated Fermi energy as a function of the gate voltage, and Fig. S4b shows the numerical derivative of the Fermi energy, $dE_F/dV_g$. As observed, at the threshold voltage $V_{th}$ the variation of the Fermi energy with the gate voltage is negligible compared with the thermal energy, $k_B T = 25$ meV. However, when the gate voltage is around 50 V below the threshold voltage, i.e. when the Fermi energy is $E_F \sim -2 k_B T$, the conduction band is only populated by thermally excited electrons, and a much higher increase in the Fermi energy is required to increase the number of charge carriers by the same amount. In consequence, in this voltage range, the variation of the Fermi energy with the gate voltage is comparable with $k_B T$.

**References**

[S1]  M. M. Furchi, D. K. Polyushkin, A. Pospischil, T. Mueller, *Nano Lett.* **2014**, 22.

21